\documentclass{aa}  

\usepackage{psfig}
\usepackage{natbib}
\usepackage[english]{babel}  
\usepackage{graphicx}
\usepackage{txfonts}
\begin{document}

   \title{Modeling the shock-cloud interaction in SN 1006: unveiling the origin of nonthermal X-ray and $\gamma-$ray emission}


   \author{M. Miceli
          \inst{1,2}
          \and
          S. Orlando\inst{2}
          \and
          V. Pereira\inst{3}
           \and
          F. Acero\inst{4}
           \and
          S. Katsuda\inst{5}
           \and
          A. Decourchelle\inst{6}
          \and
          F. P. Winkler\inst{7}
           \and
           R. Bonito\inst{1,2}
           \and
          F. Reale\inst{1,2}
           \and
          G. Peres\inst{1,2}
           \and
          J. Li\inst{8}
          \and
          G. Dubner\inst{9}
          }

   \institute{Dipartimento di Fisica \& Chimica, Universit\`a di Palermo, Piazza del Parlamento 1, 90134 Palermo, Italy\\
              \email{miceli@astropa.unipa.it}
         \and
     INAF-Osservatorio Astronomico di Palermo, Piazza del Parlamento 1, 90134 Palermo, Italy
         \and    
     Dpto. de Astrof\'isica y CC de la Atm\'osfera, Universidad Complutense de Madrid, E-28040 Madrid, Spain
        \and
     Laboratoire AIM, CEA-IRFU/CNRS/Universite Paris Diderot, Service d’Astrophysique, CEA Saclay, F-91191, Gif-sur-Yvette, France
        \and
     Department of Physics, Faculty of Science \& Engineering, Chuo University, 1-13-27 Kasuga, Bunkyo, Tokyo 112-8551, Japan
     \and
     Service d'Astrophysique, CEA Saclay, F-91191 Gif-sur-Yvette Cedex, France   
        \and
        Department of Physics, Middlebury College, Middlebury, VT 05753, USA
        \and
        Department of Astronomy, University of Michigan, 311 West Hall, 1085 S. University Ave, Ann Arbor, MI 48109-1107, USA
        \and
     Instituto de Astronom\'ia y F\'isica del Espacio (IAFE), UBA-CONICET, CC 67, Suc. 28, 1428 Buenos Aires, Argentina
     }

   \date{}

 
  \abstract
   {The supernova remnant SN~1006 is a source of high-energy particles and its southwestern limb is interacting with a dense ambient cloud, thus being a promising region for $\gamma-$ray hadronic emission.}
   {We aim at describing the physics and the nonthermal emission associated with the shock-cloud interaction to derive the physical parameters of the cloud (poorly constrained by the data analysis), to ascertain  the origin of the observed spatial variations in the spectral properties of the X-ray synchrotron emission, and to predict spectral and morphological features of the resulting $\gamma-$ray emission.}
   {We performed 3-D magnetohydrodynamic simulations modeling the evolution of SN 1006 and its interaction with the ambient cloud, and explored different model setups. By applying the REMLIGHT code on the model results, we synthesized the synchrotron X-ray emission, and compared it with actual observations, to constrain the parameters of the model. We also synthesized the leptonic and hadronic $\gamma-$ray emission from the models, deriving constraints on the energy content of the hadrons accelerated at the southwestern limb.}
   {We found that the impact of the SN 1006 shock front with a uniform cloud with density $0.5$ cm$^{-3}$ can explain the observed morphology, the azimuthal variations of the cutoff frequency of the X-ray synchrotron emission, and the shock proper motion in the interaction region. Our results show that the current upper limit for the total hadronic energy in the southwestern limb is $2.5\times10^{49}$ erg.}
   {}


   \maketitle
%

\section{Introduction}

The shock fronts of supernova remnants (SNRs) are well known sites of efficient electron acceleration, as shown by the ubiquitous presence of synchrotron radio shells (tracing the presence of GeV electrons) in most galactic SNRs \citep{gre09} and by the detection of synchrotron X-ray emission in young SNRs \citep{rey08,vin12}, proving that in these sources the electron energy can reach values of the order of 10 TeV. 

SN 1006 is an ideal target to study particle acceleration and it has been widely observed by the current generation of X-ray telescopes, specially through a dedicated \emph{XMM-Newton} Large Programme of observations ($XMM-LP$, PI: A. Decourchelle, exposure time $t_{exp}\sim700$ ks); a Chandra Large Project ($Chandra-LP$, PI: F. Winkler, $t_{exp}\sim670$ ks); and several $Suzaku$ observations. Despite its age, the remnant is dynamically young, because it evolves in a tenuous environment $\sim550$ pc above the galactic plane (assuming a distance of 2.2 kpc, \citealt{wgl03}), and its shock velocity exceeds $5000$ km$/$s (\citealt{kpl09,klp13,wwr14}).  The analysis of the deep X-ray observations of the $XMM-LP$ revealed that the ambient density is $n_{ISM}\sim 0.035$ cm$^{-3}$ in the southeastern limb \citep{mbd12} and similar estimates have been obtained within the $Chandra-LP$ \citep{wwr14}.
The bilateral morphology of the nonthermal emission reflects highly efficient particle acceleration in the radio, X-ray, and $\gamma-$ray bright northeastern and southwestern limbs and regions with less efficient particle acceleration in the northwestern and southeastern thermal limbs.
The spatial distribution of the thermal emission has shown inhomogeneities in the physical and chemical properties of the plasma (\citealt{uyk13,wwr14,ldm15}) and has provided important insight on the shock-heating mechanism \citep{bvm13}.
The nonthermal emission of SN 1006 presents significant variations in the cutoff energy of the synchrotron emission, $h\nu_{cut}$, (\citealt{rbd04,mbi09,kpm10}). Also, the shape of the cutoff in the X-ray spectra of the nonthermal limbs reveals that the maximum energy of the electrons is limited by their radiative losses \citep{mbd13,mbd14}. 
Therefore hadrons, that do not undergo significant radiative losses, may be, in principle, accelerated up to higher energies.
Indeed, effects of shock modification induced by the back-reaction of energetic hadrons have been observed in SN 1006, through variations of the shock compression ratio in the southeastern limb \citep{mbd12}, and amplification of the magnetic field (e.g., \citealt{rkr14} and references therein).

\citet{mad14} (hereafter Paper I) studied a sharp indentation in the southwestern shock front of SN 1006 (visible in X-rays and in the radio band), finding several signatures for a shock-cloud interaction. The indentation corresponds to the position of an HI cloud having the same velocity as the northwestern cloud, which is interacting with SN 1006, and the variations of the $N_H$ derived from the X-ray spectra are consistent with those obtained from the HI observations. A clear proof of the shock-cloud interaction is provided by the azimuthal profile of the synchrotron cutoff energy, which is significantly lower in the indentation than in adjacent regions. This is because  the shock is slowed down by the dense cloud and $h\nu_{cut}$ decreases with the square of the shock speed, $v_s$, in the loss-limited scenario \citep{za07}.

The unique combination of efficient particle acceleration and high target density (i.e. the cloud) makes the southwestern limb of SN 1006 a promising region for $\gamma-$ray hadronic emission (i.~e., proton-proton interactions with $\pi^0$ production and subsequent decay).
However, Paper I shows that the cutoff frequency in the interaction region is reduced only by a factor $f\sim 1.7$ and this would suggest that the cloud density, $n_{cl}$ is higher than $n_{ISM}$ by the same factor\footnote{Because $h\nu_{cut}\propto v_{s}^2\propto n^{-1}$.}, and therefore $n_{cl}<0.1$ cm$^{-3}$ (assuming $n_{ISM}\sim0.03-0.05$ cm$^{-3}$). This value is much smaller than the average density of the southwestern cloud estimated by the HI data ($\sim10$ cm$^{-3}$, see Paper I for details). This discrepancy may be due to the fact that other (non-interacting) parts of the shell may contribute to the projected synchrotron emission at the indentation, thus producing an enhancement in the measure of the cutoff energy. Also, the shock is probably interacting with the outer border of the cloud, where the density is expected to be lower. Furthermore, the estimates based on the HI data rely on assumptions about the cloud geometry and its extension along the line of sight. In any case, the data analysis does not allow us to obtain accurate constraints on the physical properties of the cloud, which are crucial to estimate the expected hadronic flux.

The HESS observations of SN 1006 \citep{aaa10} are consistent with a pure leptonic model, i.~e., inverse Compton emission (IC) from the electrons accelerated at the nonthermal limbs, in agreement with the morphology of the $\gamma-$ray emission of SN 1006, which also favors a leptonic origin \citep{pbm09}. \citet{aaa10} have shown that a mixed scenario that includes leptonic and hadronic components also provides a good fit to the $\gamma-$ray data and the model by \citet{bkv12} shows that the hadronic and leptonic components in the $\gamma-$ray emissions are of comparable strength. However, the most recent upper limits of the SN 1006 flux in the $3-30$ GeV band obtained with the $Fermi-LAT$ telescope by \citet{alr15} rule out a hadronic origin for the $\gamma-$ray emission at a $>5~\sigma$ confidence level, even in the southwestern limb. Nevertheless, given the small angular size of the shock-cloud interaction region, a hadronic contribution from the cloud can still be consistent with the data.

We present here a 3-D magneto-hydrodynamic (MHD) model of the shock-cloud interaction in SN 1006 to obtain a deeper level of diagnostics. We synthesize observables from the model that we test against actual X-ray and $\gamma-$ray observations to constrain the physical parameters of the cloud and to obtain accurate predictions on the resulting hadronic and leptonic emission. In particular, we take advantage of both the $XMM-LP$ and $Chandra-LP$ on SN 1006 to obtain tight observational constraints for our model, and of the $HESS$ and $Fermi-LAT$ observations for the $\gamma-$ray band.
The paper is organized as follows: Sect. \ref{setup} describes our model and the setups of our simulations; Sect. \ref{synth} describes the procedures followed to synthesize the emission from the model; and Sect. \ref{Results} shows the comparison between model and observations. Our conclusions are summarized in Sect. \ref{Conclusions}.


\section{MHD model and numerical setup}
\label{setup}
We performed 3-D MHD simulations describing the expansion of the whole remnant of SN 1006 in a cartesian coordinate system with the FLASH code \citep{for00}. The computational domain extends 24 pc in the x-, y-, and z-directions and we followed the evolution of the system for 1000 yr. We assumed zero-gradient (outflow) conditions at all boundaries.
The model to describe the evolution of SN 1006 is the same as that presented in \citet{obm12}; in particular, we adopted their model PL-QPAR-G1.3, with slightly revised values of the explosion energy and ambient density (see below).  

Our initial conditions were carefully tuned to reproduce SN 1006 after 1000 yr of evolution in terms of size and shock velocity. The setup consists of a spherically symmetric distribution of ejecta, centered at position $(0,0,0)$ cm, with kinetic energy $K=1.3\times10^{51}$ erg, mass $M_{ej}=1.4$ M$_{\sun}$, and initial radius $R_0=1.4\times10^{18}$ cm (corresponding to an age of $\sim10$ yr at the beginning of the simulations). The radial density profile of the ejecta follows a power-law distribution with index $n=-7$, which is typical for the outer layers of Type Ia SNe (e. g., \citealt{che82}). We compared our results with those obtained with different profiles, namely the models with an exponential profile presented in \citet{obm12} and a  step-like ejecta profile, as that adopted in \citet{mbi09}. We found that the ejecta profile influences the shape and size of the Rayleigh-Taylor and Richtmyer Meshkov instabilities that develop at the contact discontinuity between the ejecta and the shocked ISM, but the global properties of the remnant  (radius, shock velocity, distance between the forward shock and the contact discontinuity, etc.) at the age of SN 1006 do not change significantly among the inspected cases (see \citealt{obm12} for a quantitative comparison). The ejecta expand through an unperturbed magneto-static medium with density $n_{ISM}=0.035$ cm$^{-3}$ (in agreement with \citealt{mbd12}), where we placed a dense, isobaric, spherical cloud, in pressure equilibrium with the ambient medium.

\begin{figure}[htb!]
 \centerline{\includegraphics[width=\columnwidth]{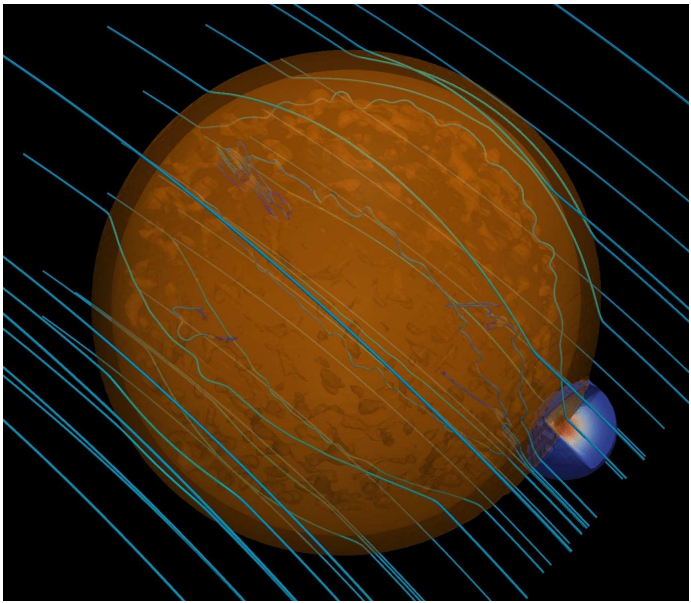}}
\caption{3D rendering of the MHD simulation RUN2\_G, describing the expansion of SN 1006 through a magnetized medium and its interaction with an ambient cloud at t = 1000 yrs (the model parameters are shown in Table \ref{tab:setups}). The image is rotated to match the actual conditions of SN 1006 and the orientation of the observations (North is up and East is to the left). The ejecta material and the shocked ambient medium are tracked with a ``solid” and a semi-transparent surface, respectively. The cutaway drawing of the southwestern cloud reveals its interior: the color code traces the density, which increases radially from $0.07$ cm$^{-3}$ (dark blue) to $10$ cm$^{-3}$ (red). The magnetic field lines are superimposed. The (projected) gradient of the magnetic field points toward southeast.}
\label{fig:run2}
\end{figure}

\begin{table*}[htb!]
\caption{Model setups}            
\label{tab:runs} 
\centering                        
\begin{tabular}{c c c c c} 
\hline\hline               
Model &  Cloud Radius  & Cloud density & Cloud position & Dipole position \\    
      & ($10^{18}$ cm) &(min-max, cm$^{-3})$  & (center, $10^{19}$ cm) &       (pc)      \\
\hline                        
RUN0\_G  & $8.1^{*}$  & $0.07-10$ &$(2.6,0.3,0.2)$ & $(0, 0, -100)$ \\   
RUN1\_G  & $8.1^{*}$  & $0.07-10$ &$(3.0,0.3,0.2)$ & $(0, 0, -150)$ \\
RUN2\_G  & $8.1^{*}$  & $0.07-10$ &$(2.8,0.3,0.4)$ & $(0, 0, -300)$ \\
RUN3\_G  & $8.1^{*}$  & $0.07-10$ &$(2.8,0.3,0.4)$ & $(0, 0, -1000)$  \\
RUN1\_UN & $6.18$ &  $0.5$    &$(2.8,0.3,0.4)$ & $(0, 0, -300.)$ \\ 
RUN2\_UN & $5.5$  &  $0.5$    &$(2.8,0.3,0.4)$ & $(0, 0, -300.)$ \\
\hline                                   
\end{tabular}
\tablefoot{$^*$ Radius at 3 sigmas of the Gaussian density distribution.}
\label{tab:setups}
\end{table*}

The average ambient magnetic field in the environment of SN 1006 is expected to be directed along the Southwest-Northeast direction (\citealt{rbd04,rhm13}) with a gradient pointing to Southeast, in the direction of the Galactic plane (\citealt{bom11} and references therein). We implemented this configuration by considering a dipole as a source of the magnetic field, as explained below.
The coordinate system was chosen so as to have the average ambient magnetic field directed along the x-axis and its gradient along the z-axis. In particular, the ambient magnetic field was generated by a dipole at position $(0,0,-d_z)$ pc, placed outside the computational domain. We explored different values of $d_z$ (see below) and in all the simulations we tuned the magnetic field strength to get $B_0\sim30$ $\mu$G in the environment of the explosion site. The magnetic field gradient leads to a variation of a factor of $\sim1.3$ over a scale of 20 pc and makes the polar caps (defined as the points where $v_s$ and $B_0$ are parallel) converge on the side in which the magnetic field is increasing (thus effecting the morphology of the nonthermal limbs, see \citealt{obr07}).

We exploited the adaptive mesh capabilities of the FLASH code by adopting up to 10 nested levels of resolution (the resolution increases by a factor of 2 at each level). The finest spatial resolution is $1.8\times10^{16}$ cm at the beginning of the simulation (i. e., 85 computational cells per initial radius of the ejecta). We adopted an automatic mesh redefinition scheme to keep the computational cost approximately constant as the blast expanded, decreasing the spatial resolution down to $2.9\times10^{17}$ cm (corresponding to 98 cells per radius of the remnant) at the end of the simulation.

We included the effects of shock modification induced by the particle acceleration as in \citet{obm12} (based on the approach by \citealt{fdb10} which relies on the Blasi model; see \citealt{bla02,bla04}). We also considered the dependence of the particle acceleration on the obliquity angle $\theta$ (i.e., the angle between $B_0$ and $v_s$) in the quasi-parallel scenario, by following Eq. 1 in \citet{obm12} with adiabatic index $\gamma_{min}=4/3$ (i. e., the shock compression ratio is $\sigma=7$ at the polar caps and $\sigma=4$ for $\theta=90^\circ$, see Sect. 2 of \citealt{obm12} for further details). 
Our model does not include the effects of magnetic field amplification due to the CRs streaming at the shock front. We then chose a relatively high value of the upstream magnetic field $B_0$ to obtain a downstream ``compressed" magnetic field of the order if $100$ $\mu$G, in agreement with observations (\citealt{mab10,rkr14}, see also Sect. \ref{Xray}).

We considered the parameters of the cloud inferred from the observations in Paper I as fiducial values and ran different simulations around these observational estimates to constrain the physical properties of the cloud. In particular, the parameters explored include: i) the position and the radius of the cloud; ii) the position of the magnetic dipole (influencing the distance of the cloud to the polar caps of the remnant); iii) the cloud density; and iv) the radial density profile of the cloud, namely a uniform density profile (RUN1,2\_UN) and a centrally peaked Gaussian profile (RUN0-3\_G). Table \ref{tab:runs} summarizes our exploration of the parameter space. 
The cloud density, which is poorly constrained by the data analysis, varies by more than two orders of magnitude in runs RUN0-3\_G. In these runs, we explored different distances of the cloud from the explosion site, therefore the SNR shock reaches different parts of the cloud at $t=1000$ yr. This allowed us to study the effects of the propagation of the transmitted shock in the tenuous exterior part of the cloud (e.~g., RUN1\_G) or in its denser core (e.~g., RUN0\_G).
The output of all our simulations were then rotated to match the actual conditions of SN 1006 and the orientation of the observations (where North is up and East is to the left). In particular, all the maps presented here are rotated by an angle $\alpha_X=15^\circ$, $\alpha_Z=8^\circ$, $\alpha_Y=37^\circ$, about the $x$, $z$, and $y$ axis, respectively (the rotations being performed in this order). With these rotations the magnetic field gradient points in the direction of the galactic plane, as suggested by \citet{bom11}. As an example, Fig. \ref{fig:run2} shows a rendering of the final stage of model RUN2\_G.  We also performed a convergence test, by increasing the resolution of model RUN2\_G by a factor of 2, and found that the results do not differ significantly from those reported in Sect. \ref{Results}.

Finally, we point out that \citet{obm12} have shown that inhomogeneities in the ejecta density profile (``clumps") can perturb the contact discontinuity and affect the distance between ejecta and shocked ISM, thus triggering the formation of shrapnel protruding even beyond the shock front (see also \citealt{mor13}). These ejecta clumps, well visible at different positions of the shell of SN 1006 (e. g., \citealt{rlh11,wwr14}) are \emph{not} present in the shock-interaction region we are modeling here (see the ``pure thermal'' image of the remnant in \citealt{mbi09}). Also,  we are interested at modeling the interaction between the outer shock front and the ambient medium and not the details of the ejecta evolution. Therefore, we did not include ejecta clumps in our model. 

\section{Synthesis of observables}
\label{synth}

To get a quantitative comparison between our MHD models and actual observations of SN 1006, we followed a forward modeling approach, by synthesizing the emission from the simulations and comparing the synthesized observables and the corresponding data.
The X-ray emission of the shock-cloud interaction region in the southwestern limb of SN 1006 is dominated by synchrotron radiation, the contribution of thermal emission being negligible (see Paper I). We then focused on the nonthermal emission and used REMLIGHT, a code developed to synthesize the synchrotron radio, X-ray, and IC $\gamma-$ray emission from MHD simulations, introduced and described in detail in \citet{opb11}. In particular, we derived the nonthermal emission in the loss-limited case (see Sect. 3.1 in \citealt{opb11}), which is appropriate for the nonthermal limbs of SN 1006 (\citealt{mbd13,mbd14}). The outputs of REMLIGHT are 3-D data cubes of i) synchrotron monochromatic X-ray emission at selected energies; ii) cutoff energy of the electron spectrum; iii) IC monochromatic emission at selected energies. 

\begin{figure*}[htb!]
 \centerline{\includegraphics[width=\linewidth]{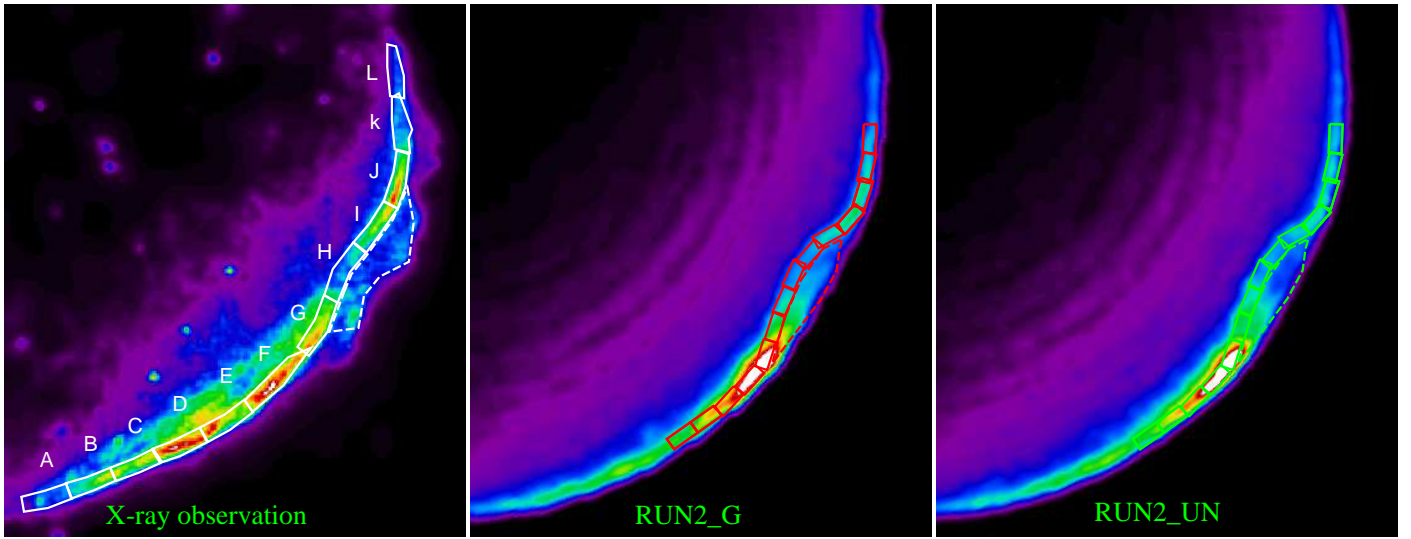}}
\caption{\emph{Left panel}: \emph{XMM-Newton} EPIC count-rate image (MOS and pn mosaic) of the southwestern limb of SN 1006 in the $2-4.5$ keV band. The regions selected for the spectral analysis of the rim performed in Paper I are superimposed. \emph{Central panel:} Synthetic synchrotron emission at 3 keV at $t=1000$ yr derived from model RUN2\_G (see Table \ref{tab:setups}). The regions selected to derive the azimuthal variation of the synchrotron cutoff energy at the shock front are superimposed. \emph{Right panel:} same as central panel for model RUN2\_UN.}  
\label{fig:Xmaps}
\end{figure*}

\subsection{Synthesis of X-ray emission}
\label{Xray}
To produce synthetic X-ray images, we calculated the synchrotron 3-D data cubes at 3 keV and summed the contribution of each cell to the emission along the line of sight. We also computed the cutoff energy $h\nu_{cut}$ of the X-ray synchrotron radiation, derived from the electron cutoff energy. In the loss limited scenario, it has the form (\citealt{za07}):
\begin{equation}
h\nu_0 = \frac{2.2 {\rm keV}}{\eta(1+\sqrt{\kappa})^2} v_{3000}^2 \frac{16}{\gamma_{s}^{2}}
\label{eq:hnu}
\end{equation}
where $\eta$ is the gyrofactor (i. e., the ratio of the electron mean free path to the gyroradius), $\kappa$ is the magnetic field compression ratio, $v_{3000}$ is the shock speed in units of $3000$ km/s, and $\gamma_s$ is the power-law index of particles accelerated at the absence of energy losses. 

We included in our model two passive tracers (defined in the whole domain) storing the time of the shock impact and the shock velocity at the shock impact for each cell, respectively. As the fluid is advected away from the shock front, this information is used to update the electron energy spectrum. The value of $h\nu_{cut}$ varies with the azimuthal angle as explained before and with the distance to the shock, as the shape of the electron energy spectrum is modified by the radiative losses. 
We imposed the maximum value of $h\nu_{cut}$ to be $h\nu^{//}_{cut}=340$ eV in the polar caps, in agreement with what is observed in the southwestern limb of SN 1006 (see Paper I). 
Then we calculated the ``effective" synchrotron cutoff energy in specific regions of the synthetic X-ray image, by performing a weighted average of the cutoff energy along the line of sight for all the cells whose projected position lies within the region, the weight being the local X-ray luminosity (which is a good proxy of the actual X-ray luminosity in the southwestern limb of SN 1006, as shown in Sect.\ref{Results}).

We point out that the parametric function adopted in our model to describe the shock obliquity dependence of the particle acceleration (described in Sect. \ref{setup}) is purely heuristic and that we are not including the effects of magnetic field amplification (MFA). The physics of MFA at the shock precursor is still debated and extremely complicated, involving resonant and non-resonant cosmic-ray driven instabilities operating at different time-scales and length-scales (see the reviews by \citealt{ber12} and \citealt{sbd12}). Different approaches have been adopted to treat this process and to couple it with the MHD evolution of a SNR (e. g., \citealt{ab06,cap12,len12,kje13}). \citet{fds14} considered two limit cases, namely, i) total damping of the amplified field in the plasma; ii) advection to the sub-shock region of the amplified field. They found that in the second case the post-shock magnetic field is one order of magnitude higher and this significantly influences the resulting synchrotron emission. In SN 1006 the situation is even more complicated, considering the (unknown) dependence of the efficiency of the particle acceleration (and then of the MFA) on the obliquity angle.  
To obtain the correct value of the post-shock magnetic field, we then tuned our ambient magnetic field so as to obtain $B\sim100$ $\mu$G at the polar caps, in agreement with the observations (\citealt{mab10,rkr14}). However, even though we get the right magnetic field at the polar caps, its azimuthal profile may not reflect the actual conditions in the remnant and, as a consequence, the large-scale morphology of our synthetic X-ray synchrotron maps may not be an accurate proxy of the actual emission. 
However, the comparison between models and X-ray observations carried out here is not sensitive to this issue considering that i) the azimuthal extension of the interacting region is relatively small and concentrated in the polar cap region and that ii) the azimuthal modulation of the synchrotron cutoff frequency in the interaction region that we chose as a benchmark for a quantitative comparison between models and X-ray observations does \emph{not} depend on $B$ in the loss-limited scenario.

\subsection{Synthesis of $\gamma-$ray emission}
\label{gamma-ray emission}

To derive the $\gamma-$ray synthetic spectral energy distribution, we adopted REMLIGHT to produce IC 3-D data cubes at selected energies (in the $10^{-4}-300$ TeV range) and summed up the contribution of all the cells in the southwestern limb. 
We only considered IC scattering of the Cosmic Microwave Background (CMB). Indeed, the expanding ejecta of young SNRs are expected to be important factories of cold dust, whose thermal infrared emission may also contribute to the total IC emission. However, a dedicated study performed with the \emph{Spitzer} telescope at $24~\mu$m and $70~\mu$m showed no emission from the southwestern limb and, in general, no evidence for dust grain formation in the SN ejecta of SN 1006 \citep{wwb13}. 
On the other hand, $Planck$ observations showed the presence of very cold gas (at a temperature $T_d\sim15$ K) in the direction of SN 1006 \citep{pla14}, but the association with the remnant cannot be proved. Therefore, we neglected any contributions from IR dust photons to the IC emission.

The recent results obtained by \citet{alr15} clearly indicate that the bulk of the SN 1006 $\gamma-$ray emission has a leptonic origin and rule out, at a confidence level $>5~\sigma$, a standard hadronic emission scenario. Therefore, we rescaled the synthetic IC SED obtained with REMLIGHT (which is in arbitrary units) to fit the TeV $HESS$ observations of the southwestern limb of SN 1006 \citep{aaa10}.

We note that, because of the interaction with a dense cloud, the hadronic emission may have a non-negligible contribution in the $Fermi-LAT$ band.
To synthesize the hadronic $\gamma-$ray emission, we considered the particle density in each cell of the shocked ambient medium (distinguishing between ISM and cloud material). We assumed two populations of high energy hadrons, both having a power law proton energy distribution (with index $\Gamma$=2.0) and let the cutoff energy $E_p^{cut1}$ of the protons accelerated at the shock transmitted inside the cloud (i. e., those colliding with the cloud material) be different from that of protons accelerated at the remnant forward shock and interacting with the tenuous ambient material ($E_p^{cut2}$). We explored different values of $E_p^{cut1,2}$ for the two populations, and of the total hadron energy in the southwestern limb, $E_p^{tot}$.
We then computed in each cell the resulting hadronic emission at selected energies in the $10^{-4}-300$ TeV range and summed up all the contributions in the southwestern limb.
The cross section for the proton-proton inelastic collision, the $\pi^0$ production, and the $\gamma-$ray emission originating from the $\pi^0$ decays were all obtained according to \citet{kab06}. 

\section{Results}
\label{Results}

\subsection{X-ray emission}
As a first step, we compared the synthetic X-ray images of SN 1006 with the actual observations in the $2-4.5$ keV energy band. The X-ray data were collected in the framework of the $XMM-LP$ and the data analysis is described in Paper I.
Among the runs listed in Table \ref{tab:setups}, only models RUN2\_G and RUN2\_UN can reproduce the observed emission of the southwestern limb of SN 1006 in terms of azimuthal extension of the interacting region and depth of the indentation, as shown in Fig. \ref{fig:Xmaps}. 
\begin{figure}[htb!]
 \centerline{\includegraphics[angle=90,width=\columnwidth]{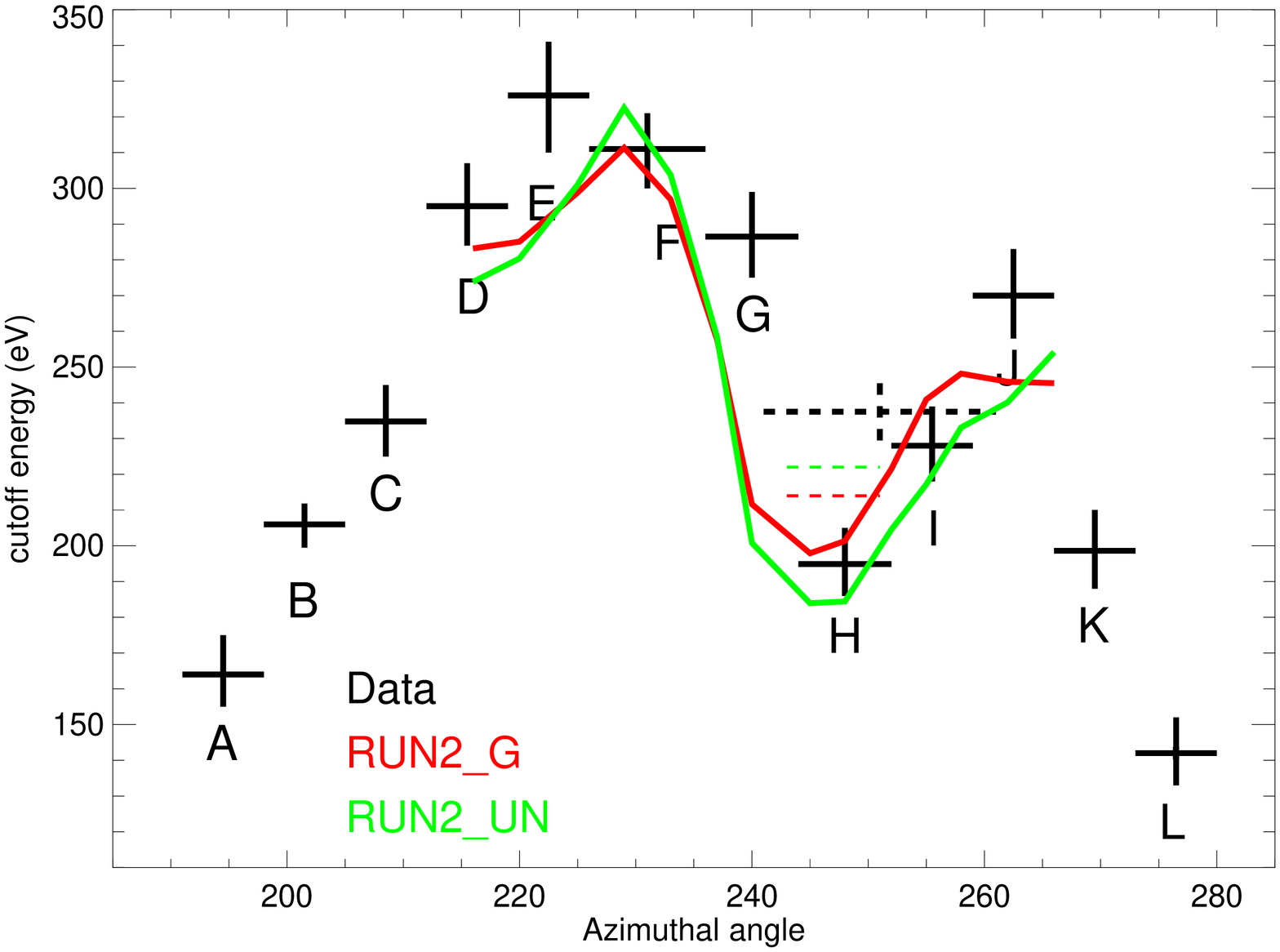}}
 \centerline{\includegraphics[angle=90,width=\columnwidth]{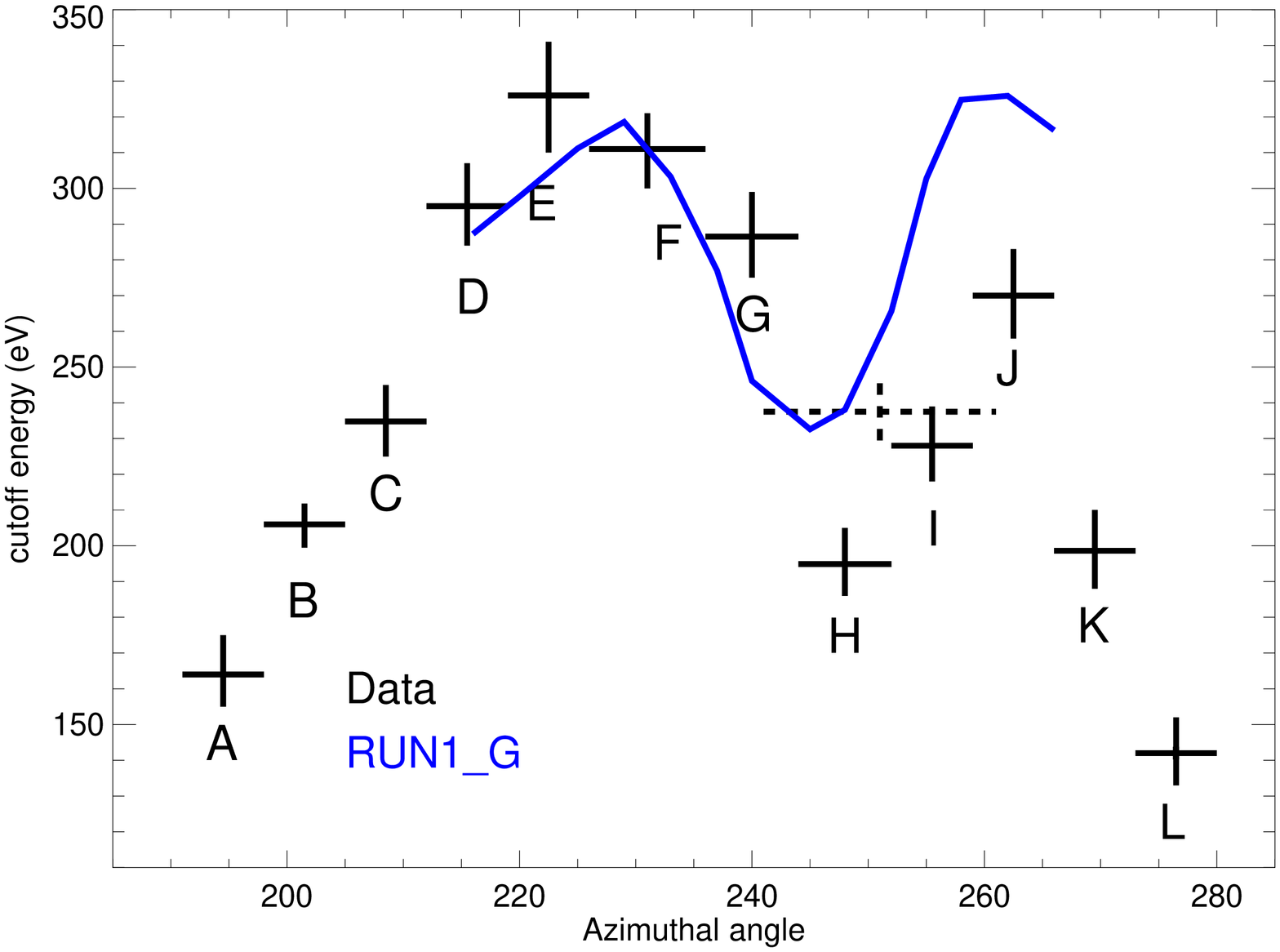}}
\caption{\emph{Upper panel:} azimuthal variations of the synchrotron cutoff energy. The black crosses show the best fit values obtained in Paper I by analyzing the X-ray spectra extracted from regions $A-L$ of Fig. \ref{fig:Xmaps}, left panel (error bars indicate the $90\%$ confidence levels). The green$/$red curves show the values derived from our MHD models RUN2\_G and RUN2\_UN, respectively, by considering the 15 regions shown in the central and right panel of Fig. \ref{fig:Xmaps}. The dashed crosses indicate the values in the ``bulge" region. \emph{Lower panel:} same as upper panel for model RUN1\_G (blue curve).}
\label{fig:cutoff}
\end{figure}

In RUN2\_G the shocked cloud has an inward increasing density profile (in the range $0.07-10$ cm$^{-3}$) and is reached by the remnant shock front at $t=640$ yr. For the first $\sim80$ yr of interaction, the transmitted shock propagates relatively fast in the outermost parts of the cloud, being only a few percent slower than the remnant forward shock that engulfs the cloud. As the transmitted shock approaches the dense core of the cloud, its speed decreases significantly, reaching a minimum value of $\sim30\%$ of the remnant shock speed at $t=1000$ yr. At this stage, the density of the cloud material immediately behind the transmitted shock front is $\sim3$ cm$^{-3}$.
In RUN2\_UN the cloud has uniform density $n_{cl}=0.5$ cm$^{-3}$ ($\sim17$ times higher than that of the surrounding medium). In this case, the cloud is slightly smaller than in RUN2\_G and is reached by the remnant forward shock at $t\sim750$ yr. At $t=1000$ yr the minimum velocity of the transmitted shock is about $40\%$ of the remnant shock speed, while the density of the shocked cloud material is $\sim1.5$ cm$^{-3}$.

In both models, the relatively low velocity of the transmitted shocks induces a drop in the synchrotron cutoff energy, which is lower by a factor $f= 8-9$ in the interaction region than in the other parts of the SN 1006 shock front. This is much higher than the observed drop of the cutoff frequency $f\sim 1.7$ derived in Paper I. However, the lateral (fast) shocks engulfing the cloud also contribute to the synchrotron emission in the (projected) interaction region and it is necessary to account for this effect to properly compare the models with the data.
We then derived maps of X-ray emission projected in the plane of the sky and selected a set of 15 regions at the shock front (shown in the central and right panels of Fig: \ref{fig:Xmaps}). We calculated the synchrotron cutoff energies in each region (as described in Sect. \ref{synth}), and compared them to those obtained in Paper I from the analysis of the X-ray spectra extracted from regions $A-L$ (shown in the left panel of \ref{fig:Xmaps}).
Upper panel of Fig. \ref{fig:cutoff} shows the observed azimuthal profile of the cutoff energy and that derived with our models. Though RUN2\_UN provides a slightly better description of the data points, both models can reproduce the observed dip in the cutoff energy and fit all the main features of the observed profile.
This result shows that we cannot discriminate between the two models on the basis of the cutoff energy variations and that it is difficult to ascertain information on the cloud structure from this parameter, given the importance of the synchrotron emission from lateral shocks in the emerging X-ray radiation.

On the other hand, we verified that the azimuthal profile of the cutoff energy in all other runs do not fit the observations. As an example, lower panel of Fig. \ref{fig:cutoff} shows the profile derived from RUN1\_G, where the cloud is placed at a larger distance to the SNR center and produces a much smaller indentation than that obtained in RUN2\_G. In this case, the dip in the cutoff energy profile is less pronounced and definitely not consistent with the data points.

We also calculated the cutoff energy in the relatively faint region ``upstream" from the indentation, marked by the dashed contours in the central and right panels of Fig. \ref{fig:Xmaps}, and compared it with that measured in the corresponding region of SN 1006 (named ``bulge" in Paper I, white dashed region in the left panel of Fig. \ref{fig:Xmaps}). The emission in this region originates in the non-interacting part of the shock front and, in model RUN2\_G, is much softer than that observed (compare the red and black dashed line in Fig. \ref{fig:cutoff}). In RUN2\_UN we achieve a better agreement between model and observations (green and black dashed line in Fig. \ref{fig:cutoff}, respectively). However, additional synchrotron emission from the bulge may arise from electrons produced by cosmic rays diffusing away from the southwestern limb in the nearby cloud and not included in our models. Also, deviations in the cloud morphology from the simple spherical$/$ellipsoidal shapes adopted in our simulations may affect the emission in this region. Therefore, we do not consider that these fits rule out model RUN2\_G. 

\begin{figure}[htb!]
  \centerline{\includegraphics[width=7cm]{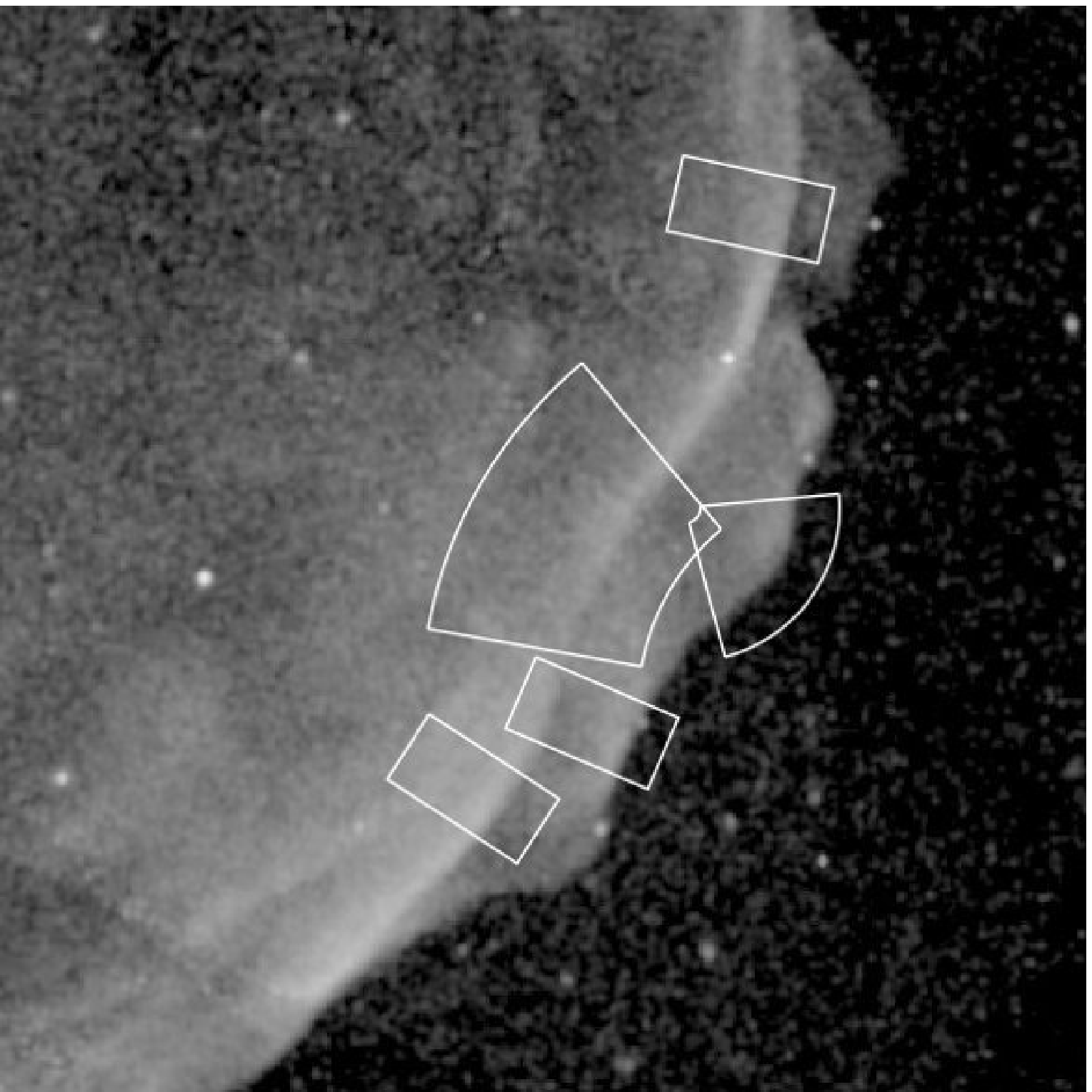}}
\centerline{\includegraphics[angle=90,width=\columnwidth]{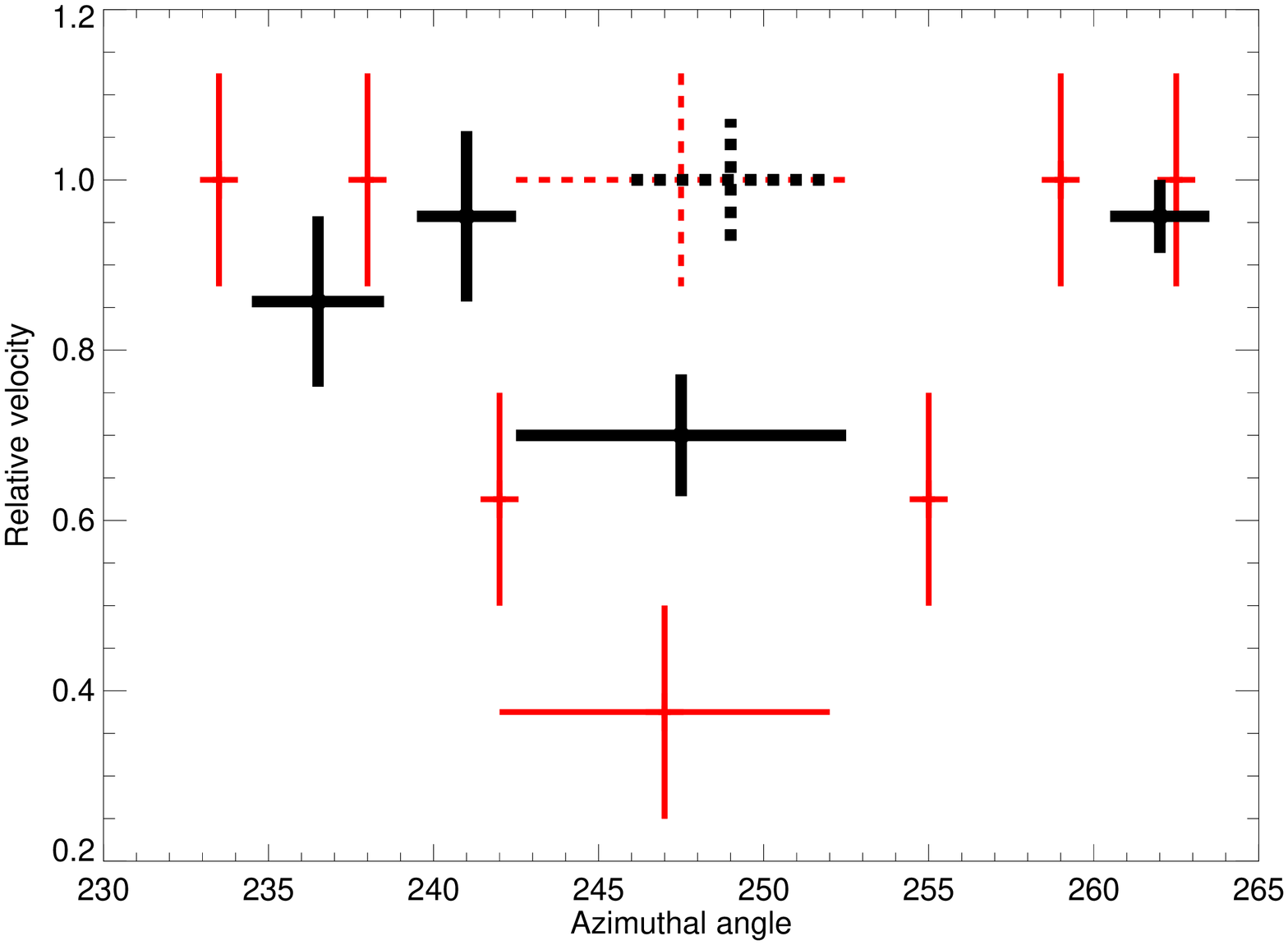}}
\centerline{\includegraphics[angle=90,width=\columnwidth]{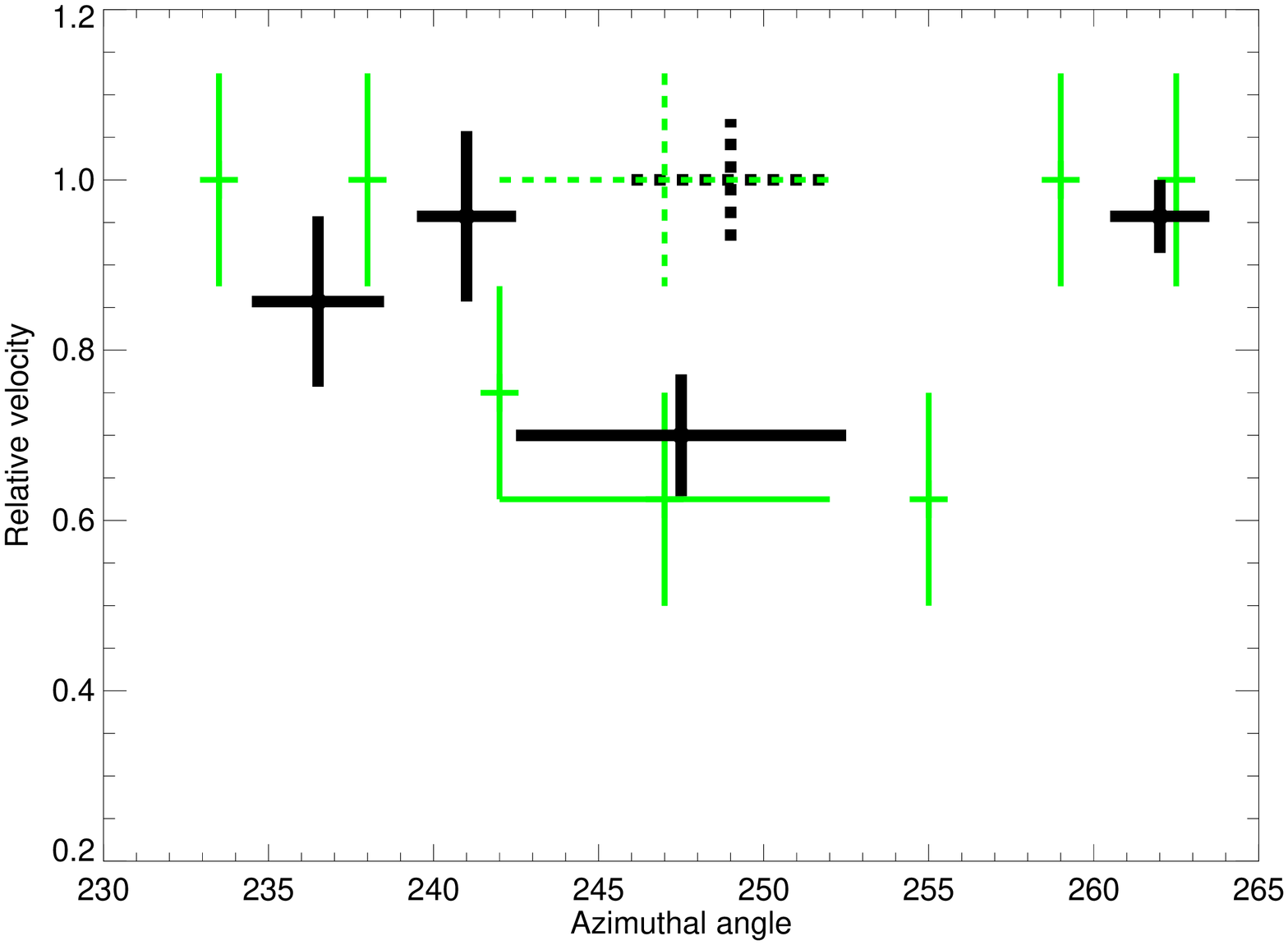}}
\caption{\emph{Upper panel:} $Chandra$ ACIS close-up view of the SN 1006 southwestern limb in the $0.5-7$ keV band. The regions selected for proper-motion measurements are shown in white. \emph{Central panel:} azimuthal variation of the proper motion (normalized to its maximum value) in the southwestern limb of SN 1006 measured with $Chandra$ (black crosses) and predicted by models RUN2\_G (red crosses). \emph{Lower panel:} same as central panel for RUN2\_UN (green crosses). The dashed crosses refer to the "bulge" region. Error bars indicate the 1-$\sigma$ confidence level for the X-ray data (black solid lines) and the sensitivity errors for the models.}
\label{fig:pm}
\end{figure}

The different densities of the shocked cloud in RUN2\_G and RUN2\_UN induce differences in the velocity of the transmitted shock which, in turn, affects the proper motion of the indentation observed in the X-ray maps. To discriminate between the two models we then measured the proper motion of the indentation in RUN2\_G and RUN2\_UN, and in the southwestern limb of SN 1006. 
\citet{wwr14} performed a systematic study of the X-ray proper motion around the whole periphery of SN 1006 by comparing the observations of the $Chandra-LP$ with previous $Chandra$ images obtained from 2003 observations and found a local minimum at the position of the indentation (azimuthal angle $\theta\sim245^\circ$). To resolve with higher resolution the azimuthal profile of the proper motion in the shock-cloud interaction region, we repeated their analysis by defining the 5 regions shown in the upper panel of Fig. \ref{fig:pm}.
As for the models, we adopted the same procedure, by selecting narrow stripes (10 pixels wide) perpendicular to the limb in the synthetic X-ray images and deriving the average radial profiles therein. We selected a much wider region in the indentation (spanning 10 degrees), as done for the real data. The error bars in the proper motion estimated by our models are sensitivity errors associated with the spatial resolution of the computational grid. Because of the limited spatial resolution of our simulations, we adopted a baseline of 40 yr.  
Lower panel of Fig. \ref{fig:pm} shows the comparison between the observed proper motion (black crosses) and that predicted by model RUN2\_G (red) and  RUN2\_UN (green). We confirm that the proper motion measured in the indentation of SN 1006 with $Chandra$ is significantly lower than that measured immediately outside the indentation and in the ``bulge". This result provides further observational evidence of the shock-cloud interaction occurring in the southwestern limb. RUN2\_G predicts a decrease in the proper motion of the indentation which is much higher than that measured. On the other hand, the proper motion predicted by RUN2\_UN is in very good agreement with the data, both for the indentation and for the ``bulge".

We conclude that RUN2\_UN is the model that best describes the X-ray emission resulting from the interaction of the SN 1006 southwestern shock front and the ambient cloud. This model, in fact, can explain: i) the morphology of the synchrotron emitting southwestern limb; ii) the azimuthal variations of the cutoff energy of the X-ray synchrotron emission; iii) the azimuthal profile of the proper motion. 

\begin{figure}[htb!]
 \centerline{\includegraphics[width=\columnwidth]{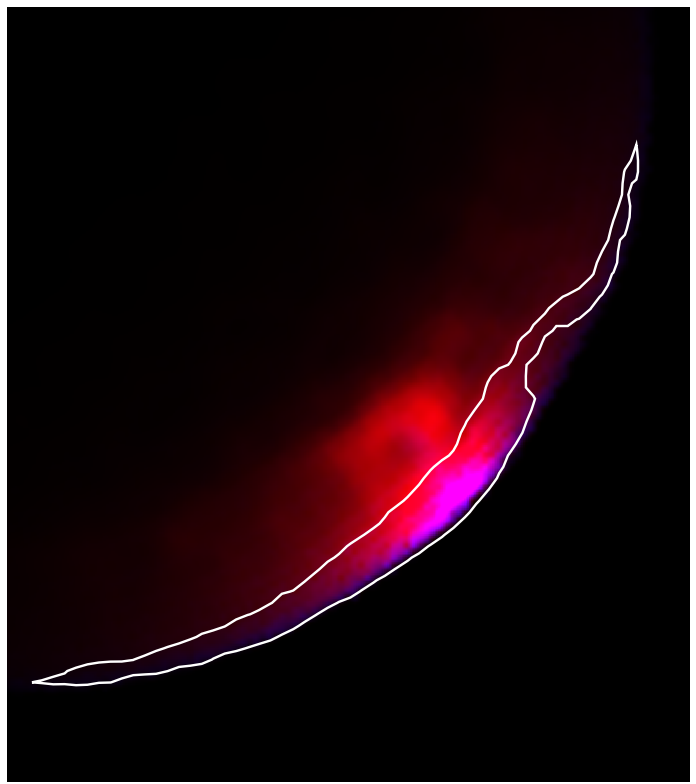}}
\caption{Synthetic inverse Compton monochromatic emission of the southwestern limb of SN 1006 at 3 GeV (in red) and 3 TeV (in blue) derived from RUN2\_UN. The contours of the X-ray emission are superimposed in white.}
\label{fig:IC}
\end{figure}

\begin{figure}[htb!]
\centerline{\includegraphics[angle=90,width=\columnwidth]{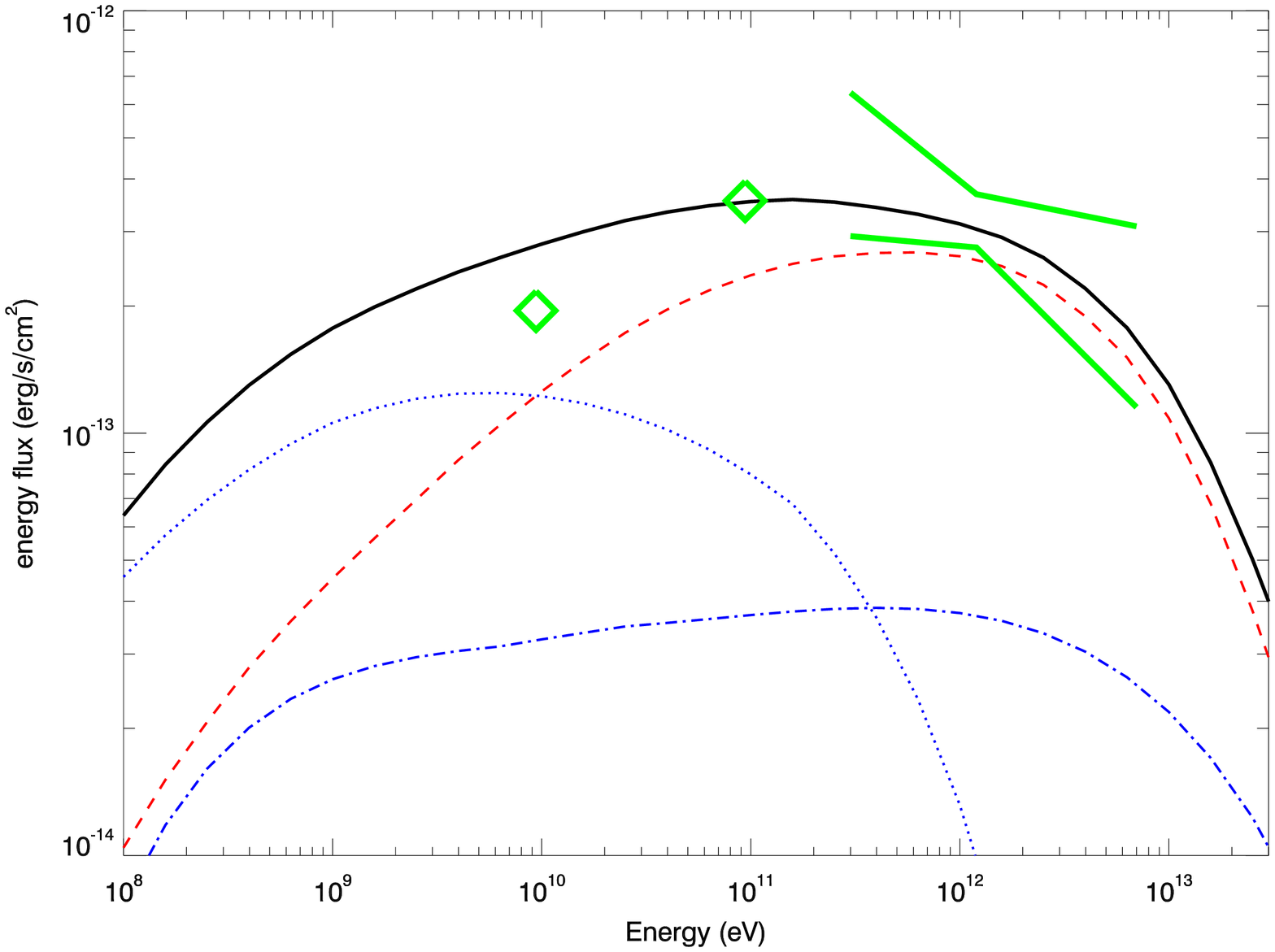}}
\centerline{\includegraphics[angle=90,width=\columnwidth]{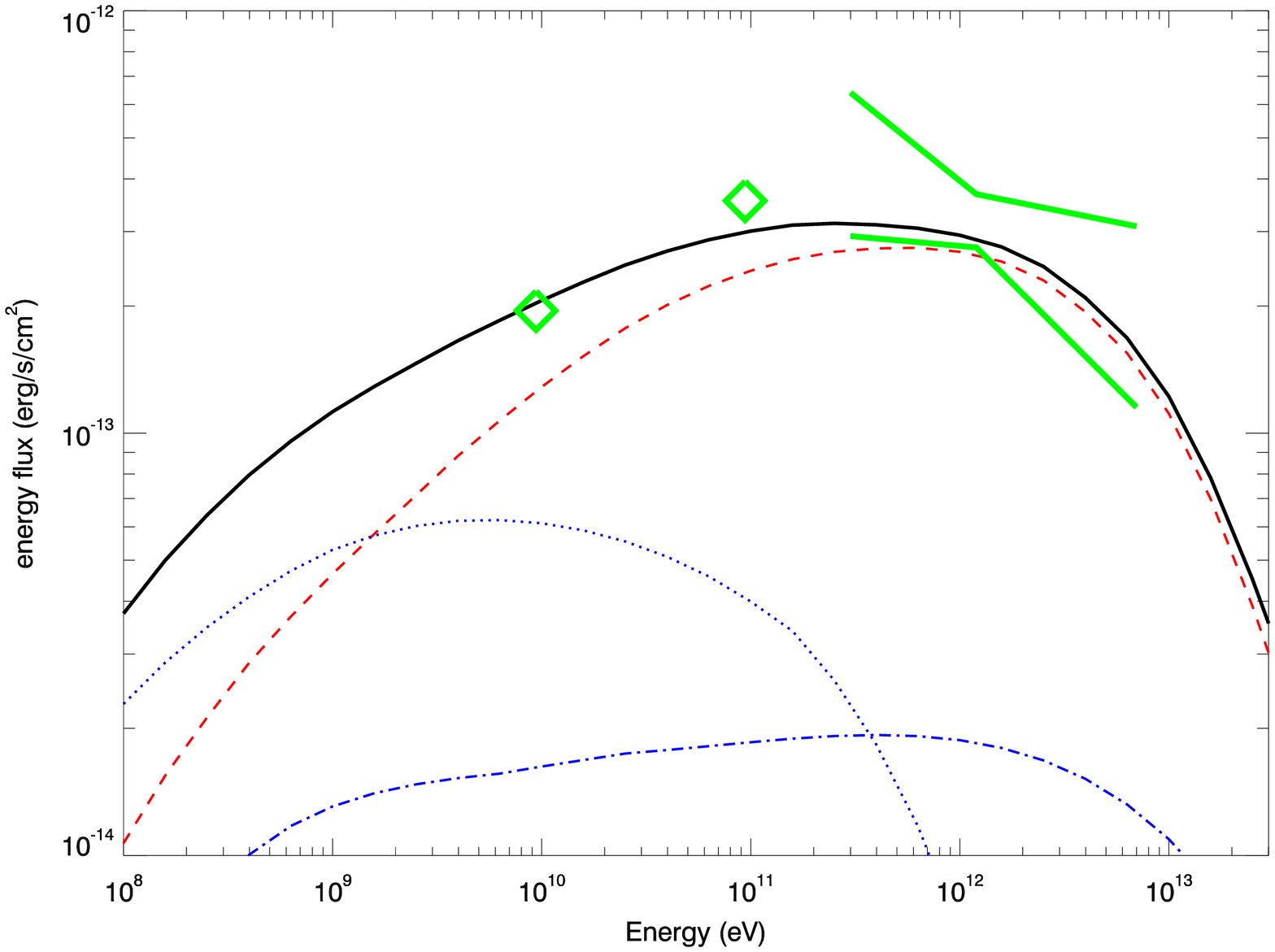}}
\centerline{\includegraphics[angle=90,width=\columnwidth]{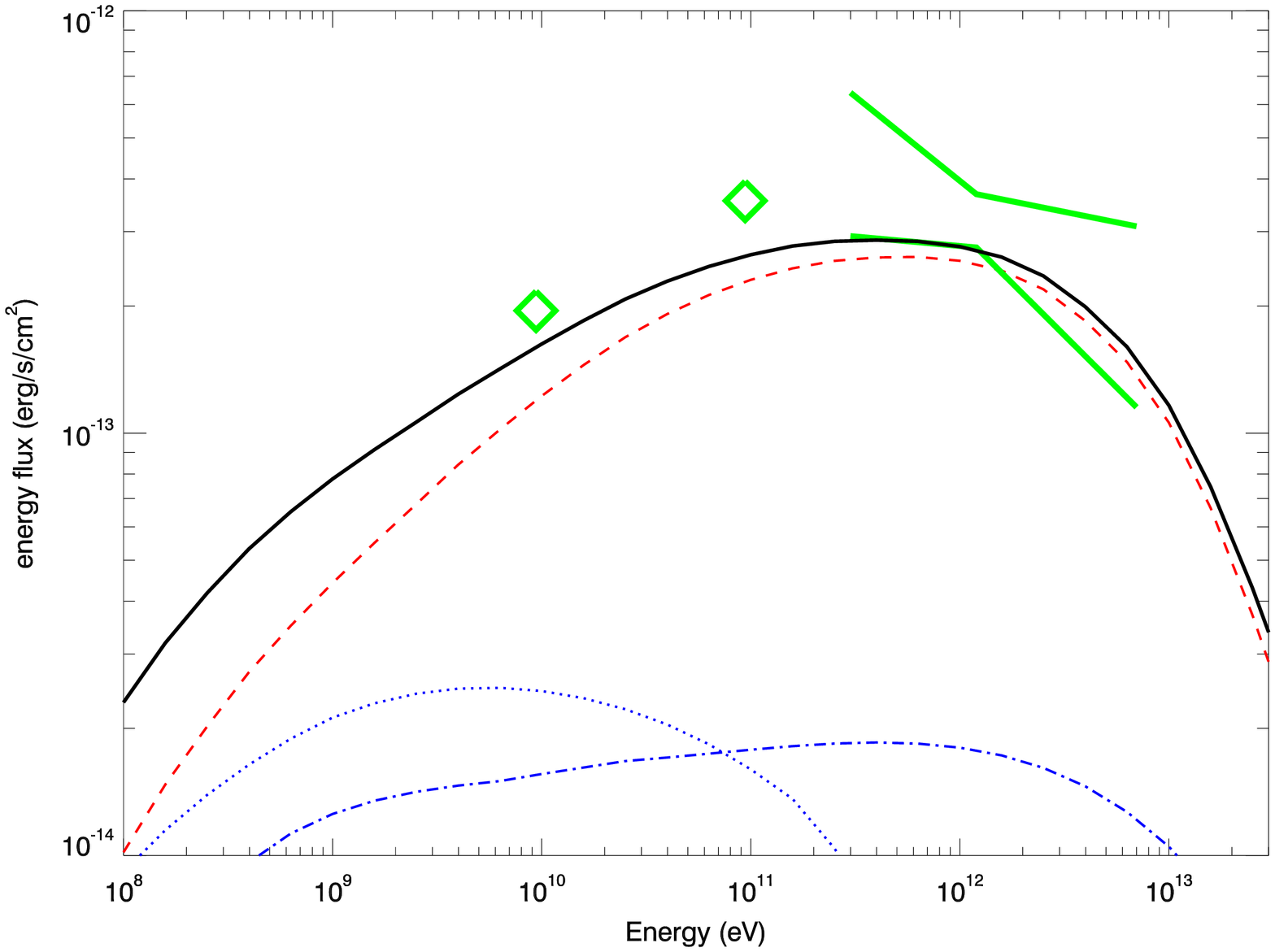}}
\caption{\emph{Upper panel:} synthetic $\gamma-$ray emission of the southwestern limb of SN 1006 obtained from RUN2\_UN by assuming a total hadronic energy $E_p^{SW}=5\times10^{49}$ erg (black solid line). The Inverse Compton contribution (dashed red line) and the hadronic contribution of the shocked cloud  (dotted blue line) and of the shocked ISM (dash-dotted blue line) are highlighted. The $\gamma-$ray spectrum observed with $HESS$ is shown in green (southwestern limb only, \citealt{aaa10}), and the $Fermi-LAT$ upper limits (at the $95\%$ confidence level) for the southwestern limb are indicated by the green diamonds (\citealt{alr15}). \emph{Central panel:} same as upper panel with $E_p^{SW}=2.5\times10^{49}$ erg. \emph{Lower panel:} same as upper panel with $E_p^{SW}=10^{49}$ erg.}
\label{fig:sedg}
\end{figure}

\begin{figure}[htb!]
 \centerline{\includegraphics[width=\columnwidth]{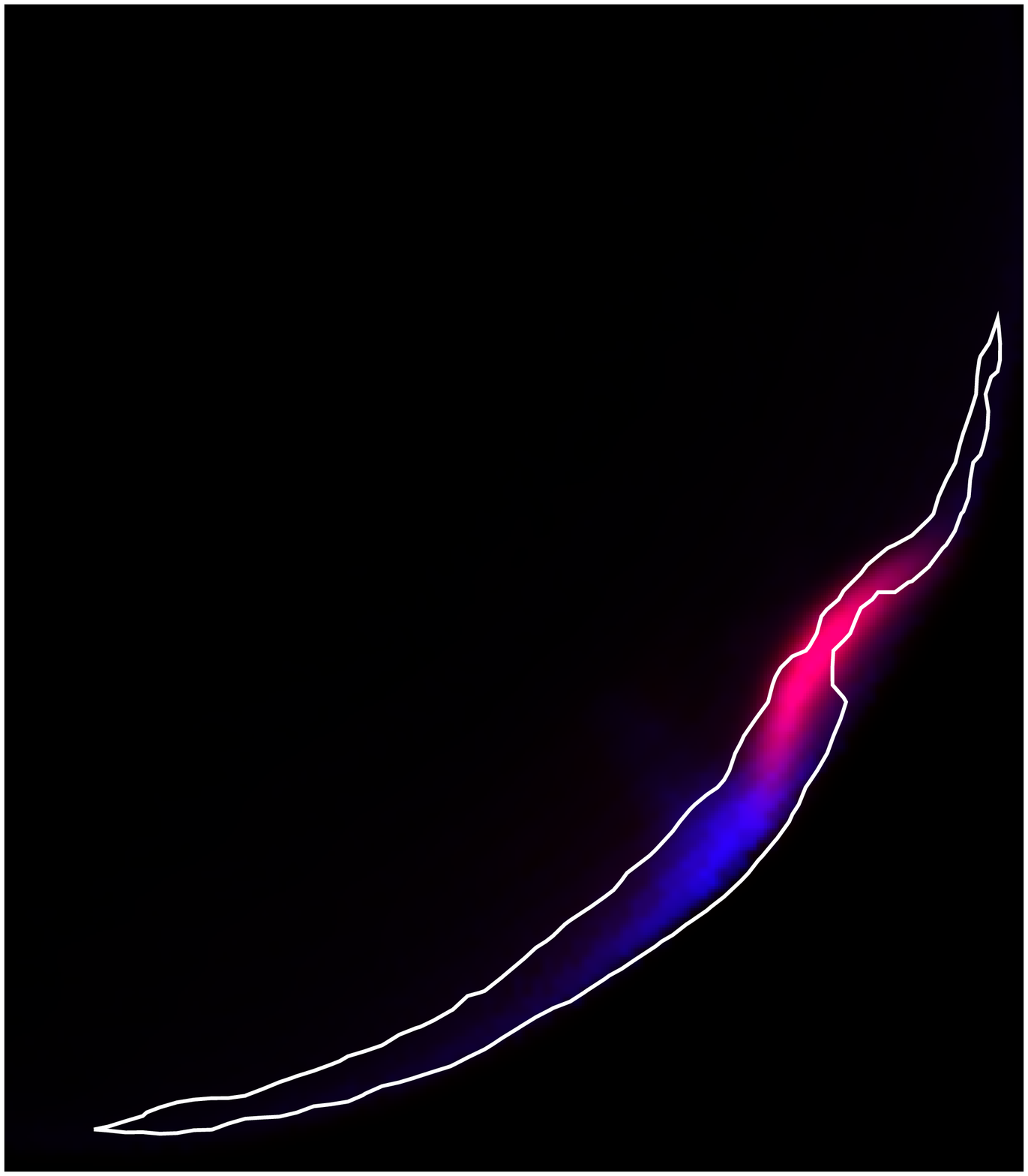}}
\caption{Synthetic $\gamma-$ray (hadronic$+$leptonic) monochromatic emission of the southwestern limb of SN 1006 at 3 GeV (in red) and 3 TeV (in blue) derived from RUN2\_UN with $E_p^{SW}=2.5\times10^{49}$ erg. The contours of the X-ray emission are superimposed in white.}
\label{fig:hf}
\end{figure}

\subsection{$\gamma-$ray emission}

Though expanding in a tenuous environment, almost six hundred pc above the Galactic plane, SN 1006 interacts with ambient interstellar clouds. A higher ambient density is observed in the northwestern rim, where the shock is slowed down by the interaction with dense material (the northwestern cloud), producing a relatively bright and sharp H$_{\alpha}$ filament and soft X-ray emission (e. g.,  \citealt{gwr02,wgl03,rks07,abd07,ldm15}). \citet{nvh13} have revealed suprathermal hadrons in the northwestern limb of SN 1006, but particle acceleration is not very efficient therein, as revealed by the very faint nonthermal emission. The southwestern limb, instead, shows both efficient particle acceleration and relatively high ambient density in the cloud, thus being a privileged site for the detection of $\gamma-$ray hadronic emission.

In Paper I we proposed a rough estimate of the expected emission on the basis of the radio and X-ray data analysis and we obtained a flux of $\sim 5\times10^{-13}$ erg cm$^{-2}$ s$^{-1}$ in the $3-30$ GeV band, by assuming a cloud density of 10 cm$^{-3}$. This flux was slightly below the $Fermi-LAT$ upper limit available.
However, the subsequent analysis of six years of $Fermi-LAT$ data performed by \citet{alr15} has recently provided more stringent constraints which seem to rule out a possible hadronic origin for the bulk of the $\gamma-$ray emission of both the southwestern and northeastern limbs of SN 1006. In particular, the upper limits (at the $95\%$ confidence level) for the flux of the southwestern limb are $\sim 1.9\times10^{-13}$ erg cm$^{-2}$ s$^{-1}$ at the median energy of the $3-30$ GeV (i. e. at $9.48$ GeV) band and $\sim 3.5\times10^{-13}$ erg cm$^{-2}$ s$^{-1}$ at $94.8$ GeV.

Here, we can refine our predictions taking advantage of the results of the MHD simulations that allowed us to tightly constrain the cloud density and the mass of the shocked cloud material. The detailed description of the targets of the proton-proton collisions leaves only the spectrum of the accelerated particle as a free parameter to derive the possible hadron emission from SN 1006, thus allowing us to ascertain some properties of the cosmic rays accelerated at the SW limb of SN 1006. In particular, we explored what values of $E_p^{tot}$ and $E_p^{cut1,2}$ are consistent with the observational constraints on the $\gamma-$ray emission.

To synthesize the $\gamma-$ray emission we then focused on our best-fit model RUN2\_UN and considered three components, namely i) the IC emission; ii) the hadronic emission originating from the impact of high-energy protons with the cloud material; and iii) the hadronic emission originating from the impact of high-energy protons with the ambient tenuous medium (see Sect. \ref{gamma-ray emission} for details). 

Fig. \ref{fig:IC} shows the synthetic IC emission at 3 GeV (in red) and 3 TeV (in blue) obtained from RUN2\_UN (a very similar map is obtained for RUN2\_G). As expected, the TeV emission, which is associated with electrons at energies $E>10$ TeV, is concentrated in the immediate post-shock region. The high energy electrons lose rapidly their energy via radiative cooling as they are advected in the post-shock region and this makes the TeV emission radially thin (and pretty similar to the X-ray synchrotron emission). Electrons with energies of a few $10^{11}$ eV, which up-scatter the CMB photons up to GeV energies, are instead present at larger distances from the shock front, thus making the GeV emission much broader in the radial direction.

As for the hadron emission, we first explored the case proposed in Paper I, with $E_p^{SW}=5\times10^{49}$ erg (i. e. a total hadronic energy $E_p^{tot}=10^{50}$ erg in the whole remnant, corresponding to $\sim10\%$ of the explosion energy), $E_p^{cut1}=3$ TeV, and $E_p^{cut2}=150$ TeV. We verified that, with this set of parameters, the resulting $\gamma-$ray emission is well above the latest $Fermi-LAT$ upper limits, as shown in the upper panel of Fig. \ref{fig:sedg}. We stress that the same happens also for RUN2\_G, where, in general, we get higher hadronic emission, given the higher average density of the shocked cloud.

We then explored different proton spectra, by reducing the total hadronic energy. Central panel of Fig. \ref{fig:sedg} shows the synthetic $\gamma-$ray emission obtained for $E_p^{SW}=2.5\times10^{49}$ erg (i. e., hadrons have drained $\sim5\%$ of the explosion energy), $E_p^{cut1}=3$ TeV, and $E_p^{cut2}=150$ TeV. With this set of parameters, the synthetic spectrum fits the observed $HESS$ spectrum at TeV energy (where the IC contribution dominates) and is consistent with the newest $Fermi-LAT$ upper limits, being at the edge of detectability in the $3-30$ GeV band. This emission, if present, will be detectable within a few more years. Lower panel of Fig. \ref{fig:sedg} shows the $\gamma-$ray emission for $E_p^{SW}=10^{49}$ erg: in this case, the $\gamma-$ray flux is well below the $Fermi-LAT$ upper limits.

\section{Summary and conclusions}
\label{Conclusions}

We performed a set of 3-D MHD simulations describing the evolution of SN 1006 and its interaction with an ambient cloud. Taking into account the estimates derived with the radio and X-ray data analysis performed in Paper I, we explored different simulation setups by modifying the properties of the ambient cloud and magnetic field.
We adopted a forward modeling approach by synthesizing observables from the simulations and comparing them against actual data.

We first focused on the X-ray morphology and identified two possible configurations, namely RUN2\_G and RUN2\_UN, whose synchrotron X-ray maps exhibit a net indentation in the X-ray southwestern limb, corresponding to the shock-cloud interaction region, which is in very good agreement with the X-ray observations. We also verified that both these runs can explain the observed azimuthal profile of the synchrotron cutoff energy. 
In RUN\_2G, the cloud has a radius $R_{cl}=8.1\times10^{18}$ cm) and an inward increasing density profile (spanning from $0.07$ cm$^{-3}$ up to $10$ cm$^{-3}$). The downstream density is $\sim3$ cm$^{-3}$ at the stage of evolution corresponding to the current conditions of SN 1006. In RUN2\_UN the cloud is slightly smaller ($R_{cl}=6.18\times10^{18}$ cm), but has a uniform density $n_{cl}=0.5$ cm$^{-3}$.
The 3-D modeling and the synthesis of the observables allowed us to explain the relatively low dip observed in the cutoff energy azimuthal profile, which appeared to be much lower than that expected by considering the cloud/ISM density contrast. In particular, we showed that the drop in the cutoff energy cannot be used as a reliable density contrast indicator because it does not trace the interacting region only, being also affected by the synchrotron emission originating from lateral shocks and integrated along the line of sight.
To discriminate between the two models we then measured the proper motion of the X-ray emitting southwestern limb and found that only RUN2\_UN provides results in agreement with that measured with the most recent $Chandra$ observations. 
Therefore, the quantitative comparison between our models and X-ray data allowed us to explain all the puzzling features observed in the southwestern limb of SN 1006 (morphology, spectral inhomogeneities of the synchrotron emission, and azimuthal variations of the proper motion). 

We tightly constrained the cloud density and the mass of the shocked cloud. In particular, we found that the pre-shock density of the cloud is $0.5$ cm$^{-3}$. In Paper I we estimated a higher cloud density from the HI data ($\sim10$ cm$^{-3}$). However, we point out that the density estimate derived in Paper I strongly relies on (arbitrary) assumption about the cloud geometry and its extension along the line of sight.
Furthermore, the bulk of the HI cloud is still unshocked, so its density may be higher than that of the interacting part. We also notice that our density estimate is pretty similar to that derived for the northwestern cloud (\citealt{rks07,abd07}), which is probably physically connected with the southwestern cloud, as suggested in Paper I. Future observations of the $^{12}$CO emission (in the $2-1$ line transition) are planned in the direction of the interacting HI gas to resolve the cloud internal structure and identify the scale of gas clumpiness to compare with the predicted physical conditions.

The determination of the cloud properties obtained with our model is crucial because the hadronic $\gamma-$ray emission depends on the spectrum of the accelerated cosmic rays and on the ambient density. Therefore the knowledge of the physical parameters of the shocked cloud allowed us to ascertain information on the hadronic acceleration in the southwestern limb. In particular, we found that if we assume that the total cosmic ray energy is $10\%$ of the explosion energy (i. e,. the hadronic energy in the southwestern limb is $5\times10^{49}$ erg), the $\gamma-$ray emission from the shocked cloud is expected to be much higher than the current $Fermi-LAT$ upper limit (as shown in Fig. \ref{fig:sedg}). We therefore have to exclude such an efficient energy drain, unless we impose that the cutoff energy of the proton spectrum in the shock-cloud interaction region is much lower than our assumed value $E_p^{cut1}=3$ TeV. In this case, the ``hadronic bump'' would be centered at much lower energies, well below the $Fermi-LAT$ sensitivity band. 

However, if the proton cutoff energy is of the order of a few TeV we derive that the upper limit for the hadronic energy in the southwestern limb is $2.5\times10^{49}$ erg (indicating a total hadronic energy of the order of $\sim5\%$ of the explosion energy). For such an energy value we expect to observe a significant hadronic emission originating in the shock-cloud interaction region and detectable with $Fermi-LAT$ within a few years. This emission will be concentrated in the shock-cloud interaction, as shown in Fig. \ref{fig:hf}. A non-detection would imply a much lower energy for the cosmic rays accelerated at the shock front.

\begin{acknowledgements}
We thank the anonymous referee for their comments and suggestions. The software used in this work was in part developed by the DOE-supported ASC / Alliance Center for Astrophysical Thermonuclear Flashes at the University of Chicago. We acknowledge the HPC facility at CINECA, Italy, and the HPC facility SCAN of the INAF – Osservatorio Astronomico di Palermo for the availability of high performance computing resources and support. This paper was partially funded by the PRIN INAF 2014 grant “Filling the gap between supernova explosions and their remnants through magnetohydrodynamic modeling and high performance computing”. MM thanks E. Amato for interesting discussions and suggestions. VP acknowledges the support of the Spanish Ministerio de Econom\'ia y Competitividad through grant AYA2011-29754-C03. PFW acknowledges the support of the National Aeronautics and Space Administration through Chandra Grant Number GO2-13066. The work of GD is funded by PICT-ANPCyT 0571/11 and PIP-CONICET 0736/11 of Argentina.
\end{acknowledgements}

 \bibliographystyle{aa}

\begin{thebibliography}{50}
\expandafter\ifx\csname natexlab\endcsname\relax\def\natexlab#1{#1}\fi

\bibitem[{{Acero} {et~al.}(2010){Acero}, {Aharonian}, {Akhperjanian}, {Anton},
  {Barres de Almeida}, {Bazer-Bachi}, {Becherini}, {Behera}, {Beilicke},
  {Bernl{\"o}hr}, {Bochow}, {Boisson}, {Bolmont}, {Borrel}, {Brucker}, {Brun},
  {Brun}, {B{\"u}hler}, {Bulik}, {B{\"u}sching}, {Boutelier}, {Chadwick},
  {Charbonnier}, {Chaves}, {Cheesebrough}, {Conrad}, {Chounet}, {Clapson},
  {Coignet}, {Dalton}, {Daniel}, {Davids}, {Degrange}, {Deil}, {Dickinson},
  {Djannati-Ata{\"i}}, {Domainko}, {O'C.~Drury}, {Dubois}, {Dubus}, {Dyks},
  {Dyrda}, {Egberts}, {Eger}, {Espigat}, {Fallon}, {Farnier}, {Fegan},
  {Feinstein}, {Fiasson}, {F{\"o}rster}, {Fontaine}, {F{\"u}{\ss}ling},
  {Gabici}, {Gallant}, {G{\'e}rard}, {Gerbig}, {Giebels}, {Glicenstein},
  {Gl{\"u}ck}, {Goret}, {G{\"o}ring}, {Hauser}, {Hauser}, {Heinz},
  {Heinzelmann}, {Henri}, {Hermann}, {Hinton}, {Hoffmann}, {Hofmann},
  {Hofverberg}, {Holleran}, {Hoppe}, {Horns}, {Jacholkowska}, {de Jager},
  {Jahn}, {Jung}, {Katarzy{\'n}ski}, {Katz}, {Kaufmann}, {Kerschhaggl},
  {Khangulyan}, {Kh{\'e}lifi}, {Keogh}, {Klochkov}, {Klu{\'z}niak}, {Kneiske},
  {Komin}, {Kosack}, {Kossakowski}, {Lamanna}, {Lemoine-Goumard}, {Lenain},
  {Lohse}, {Marandon}, {Marcowith}, {Masbou}, {Maurin}, {McComb}, {Medina},
  {M{\'e}hault}, {Moderski}, {Moulin}, {Naumann-Godo}, {de Naurois}, {Nedbal},
  {Nekrassov}, {Nicholas}, {Niemiec}, {Nolan}, {Ohm}, {Olive}, {de O{\~n}a
  Wilhelmi}, {Orford}, {Ostrowski}, {Panter}, {Paz Arribas}, {Pedaletti},
  {Pelletier}, {Petrucci}, {Pita}, {P{\"u}hlhofer}, {Punch}, {Quirrenbach},
  {Raubenheimer}, {Raue}, {Rayner}, {Reimer}, {Renaud}, {de Los Reyes},
  {Rieger}, {Ripken}, {Rob}, {Rosier-Lees}, {Rowell}, {Rudak}, {Rulten},
  {Ruppel}, {Ryde}, {Sahakian}, {Santangelo}, {Schlickeiser}, {Sch{\"o}ck},
  {Sch{\"o}nwald}, {Schwanke}, {Schwarzburg}, {Schwemmer}, {Shalchi}, {Sushch},
  {Sikora}, {Skilton}, {Sol}, {Stawarz}, {Steenkamp}, {Stegmann}, {Stinzing},
  {Superina}, {Szostek}, {Tam}, {Tavernet}, {Terrier}, {Tibolla}, {Tluczykont},
  {van Eldik}, {Vasileiadis}, {Venter}, {Venter}, {Vialle}, {Vincent}, {Vink},
  {Vivier}, {V{\"o}lk}, {Volpe}, {Vorobiov}, {Wagner}, {Ward}, {Zdziarski},
  {Zech}, \& {H.E.S.S.~Collaboration}}]{aaa10}
{Acero}, F., {Aharonian}, F., {Akhperjanian}, A.~G., {et~al.} 2010, \aap, 516,
  A62

\bibitem[{{Acero} {et~al.}(2007){Acero}, {Ballet}, \& {Decourchelle}}]{abd07}
{Acero}, F., {Ballet}, J., \& {Decourchelle}, A. 2007, \aap, 475, 883

\bibitem[{{Acero} {et~al.}(2015){Acero}, {Lemoine-Goumard}, {Renaud}, {Ballet},
  {Hewitt}, {Rousseau}, \& {Tanaka}}]{alr15}
{Acero}, F., {Lemoine-Goumard}, M., {Renaud}, M., {et~al.} 2015, \aap, 580, A74

\bibitem[{{Amato} \& {Blasi}(2006)}]{ab06}
{Amato}, E. \& {Blasi}, P. 2006, \mnras, 371, 1251

\bibitem[{{Berezhko} {et~al.}(2012){Berezhko}, {Ksenofontov}, \&
  {V{\"o}lk}}]{bkv12}
{Berezhko}, E.~G., {Ksenofontov}, L.~T., \& {V{\"o}lk}, H.~J. 2012, \apj, 759,
  12

\bibitem[{{Blasi}(2002)}]{bla02}
{Blasi}, P. 2002, Astroparticle Physics, 16, 429

\bibitem[{{Blasi}(2004)}]{bla04}
{Blasi}, P. 2004, Astroparticle Physics, 21, 45

\bibitem[{{Bocchino} {et~al.}(2011){Bocchino}, {Orlando}, {Miceli}, \&
  {Petruk}}]{bom11}
{Bocchino}, F., {Orlando}, S., {Miceli}, M., \& {Petruk}, O. 2011, \aap, 531,
  A129

\bibitem[{{Broersen} {et~al.}(2013){Broersen}, {Vink}, {Miceli}, {Bocchino},
  {Maurin}, \& {Decourchelle}}]{bvm13}
{Broersen}, S., {Vink}, J., {Miceli}, M., {et~al.} 2013, \aap, 552, A9

\bibitem[{{Bykov} {et~al.}(2012){Bykov}, {Ellison}, \& {Renaud}}]{ber12}
{Bykov}, A.~M., {Ellison}, D.~C., \& {Renaud}, M. 2012, \ssr, 166, 71

\bibitem[{{Caprioli}(2012)}]{cap12}
{Caprioli}, D. 2012, \jcap, 7, 038

\bibitem[{{Chevalier}(1982)}]{che82}
{Chevalier}, R.~A. 1982, \apj, 258, 790

\bibitem[{{Ferrand} {et~al.}(2010){Ferrand}, {Decourchelle}, {Ballet},
  {Teyssier}, \& {Fraschetti}}]{fdb10}
{Ferrand}, G., {Decourchelle}, A., {Ballet}, J., {Teyssier}, R., \&
  {Fraschetti}, F. 2010, \aap, 509, L10+

\bibitem[{{Ferrand} {et~al.}(2014){Ferrand}, {Decourchelle}, \&
  {Safi-Harb}}]{fds14}
{Ferrand}, G., {Decourchelle}, A., \& {Safi-Harb}, S. 2014, \apj, 789, 49

\bibitem[{{Fryxell} {et~al.}(2000){Fryxell}, {Olson}, {Ricker}, {Timmes},
  {Zingale}, {Lamb}, {MacNeice}, {Rosner}, {Truran}, \& {Tufo}}]{for00}
{Fryxell}, B., {Olson}, K., {Ricker}, P., {et~al.} 2000, \apjs, 131, 273

\bibitem[{{Ghavamian} {et~al.}(2002){Ghavamian}, {Winkler}, {Raymond}, \&
  {Long}}]{gwr02}
{Ghavamian}, P., {Winkler}, P.~F., {Raymond}, J.~C., \& {Long}, K.~S. 2002,
  \apj, 572, 888

\bibitem[{{Green}(2009)}]{gre09}
{Green}, D.~A. 2009, Bulletin of the Astronomical Society of India, 37, 45

\bibitem[{{Kang} {et~al.}(2013){Kang}, {Jones}, \& {Edmon}}]{kje13}
{Kang}, H., {Jones}, T.~W., \& {Edmon}, P.~P. 2013, \apj, 777, 25

\bibitem[{{Katsuda} {et~al.}(2013){Katsuda}, {Long}, {Petre}, {Reynolds},
  {Williams}, \& {Winkler}}]{klp13}
{Katsuda}, S., {Long}, K.~S., {Petre}, R., {et~al.} 2013, \apj, 763, 85

\bibitem[{{Katsuda} {et~al.}(2009){Katsuda}, {Petre}, {Long}, {Reynolds},
  {Winkler}, {Mori}, \& {Tsunemi}}]{kpl09}
{Katsuda}, S., {Petre}, R., {Long}, K.~S., {et~al.} 2009, \apj, 692, L105

\bibitem[{{Katsuda} {et~al.}(2010){Katsuda}, {Petre}, {Mori}, {Reynolds},
  {Long}, {Winkler}, \& {Tsunemi}}]{kpm10}
{Katsuda}, S., {Petre}, R., {Mori}, K., {et~al.} 2010, \apj, 723, 383

\bibitem[{{Kelner} {et~al.}(2006){Kelner}, {Aharonian}, \& {Bugayov}}]{kab06}
{Kelner}, S.~R., {Aharonian}, F.~A., \& {Bugayov}, V.~V. 2006, \prd, 74, 034018

\bibitem[{{Lee} {et~al.}(2012){Lee}, {Ellison}, \& {Nagataki}}]{len12}
{Lee}, S.-H., {Ellison}, D.~C., \& {Nagataki}, S. 2012, \apj, 750, 156

\bibitem[{{Li} {et~al.}(2015){Li}, {Decourchelle}, {Miceli}, {Vink}, \&
  {Bocchino}}]{ldm15}
{Li}, J.-T., {Decourchelle}, A., {Miceli}, M., {Vink}, J., \& {Bocchino}, F.
  2015, \mnras, 453, 3953

\bibitem[{{Miceli} {et~al.}(2014{\natexlab{a}}){Miceli}, {Acero}, {Dubner},
  {Decourchelle}, {Orlando}, \& {Bocchino}}]{mad14}
{Miceli}, M., {Acero}, F., {Dubner}, G., {et~al.} 2014{\natexlab{a}}, \apjl,
  782, L33

\bibitem[{{Miceli} {et~al.}(2012){Miceli}, {Bocchino}, {Decourchelle},
  {Maurin}, {Vink}, {Orlando}, {Reale}, \& {Broersen}}]{mbd12}
{Miceli}, M., {Bocchino}, F., {Decourchelle}, A., {et~al.} 2012, \aap, 546, A66

\bibitem[{{Miceli} {et~al.}(2013{\natexlab{a}}){Miceli}, {Bocchino},
  {Decourchelle}, {Vink}, {Broersen}, \& {Orlando}}]{mbd13}
{Miceli}, M., {Bocchino}, F., {Decourchelle}, A., {et~al.} 2013{\natexlab{a}},
  \aap, 556, A80

\bibitem[{{Miceli} {et~al.}(2014{\natexlab{b}}){Miceli}, {Bocchino},
  {Decourchelle}, {Vink}, {Broersen}, \& {Orlando}}]{mbd14}
{Miceli}, M., {Bocchino}, F., {Decourchelle}, A., {et~al.} 2014{\natexlab{b}},
  Astronomische Nachrichten, 335, 252

\bibitem[{{Miceli} {et~al.}(2009){Miceli}, {Bocchino}, {Iakubovskyi},
  {Orlando}, {Telezhinsky}, {Kirsch}, {Petruk}, {Dubner}, \&
  {Castelletti}}]{mbi09}
{Miceli}, M., {Bocchino}, F., {Iakubovskyi}, D., {et~al.} 2009, \aap, 501, 239

\bibitem[{{Miceli} {et~al.}(2013{\natexlab{b}}){Miceli}, {Orlando}, {Reale},
  {Bocchino}, \& {Peres}}]{mor13}
{Miceli}, M., {Orlando}, S., {Reale}, F., {Bocchino}, F., \& {Peres}, G.
  2013{\natexlab{b}}, \mnras, 430, 2864

\bibitem[{{Morlino} {et~al.}(2010){Morlino}, {Amato}, {Blasi}, \&
  {Caprioli}}]{mab10}
{Morlino}, G., {Amato}, E., {Blasi}, P., \& {Caprioli}, D. 2010, \mnras, 405,
  L21

\bibitem[{{Nikoli{\'c}} {et~al.}(2013){Nikoli{\'c}}, {van de Ven}, {Heng},
  {Kupko}, {Husemann}, {Raymond}, {Hughes}, \& {Falc{\'o}n-Barroso}}]{nvh13}
{Nikoli{\'c}}, S., {van de Ven}, G., {Heng}, K., {et~al.} 2013, Science, 340,
  45

\bibitem[{{Orlando} {et~al.}(2012){Orlando}, {Bocchino}, {Miceli}, {Petruk}, \&
  {Pumo}}]{obm12}
{Orlando}, S., {Bocchino}, F., {Miceli}, M., {Petruk}, O., \& {Pumo}, M.~L.
  2012, \apj, 749, 156

\bibitem[{{Orlando} {et~al.}(2007){Orlando}, {Bocchino}, {Reale}, {Peres}, \&
  {Petruk}}]{obr07}
{Orlando}, S., {Bocchino}, F., {Reale}, F., {Peres}, G., \& {Petruk}, O. 2007,
  \aap, 470, 927

\bibitem[{{Orlando} {et~al.}(2011){Orlando}, {Petruk}, {Bocchino}, \&
  {Miceli}}]{opb11}
{Orlando}, S., {Petruk}, O., {Bocchino}, F., \& {Miceli}, M. 2011, \aap, 526,
  A129

\bibitem[{{Petruk} {et~al.}(2009){Petruk}, {Bocchino}, {Miceli}, {Dubner},
  {Castelletti}, {Orlando}, {Iakubovskyi}, \& {Telezhinsky}}]{pbm09}
{Petruk}, O., {Bocchino}, F., {Miceli}, M., {et~al.} 2009, \mnras, 399, 157

\bibitem[{{Planck Collaboration} {et~al.}(2014){Planck Collaboration},
  {Abergel}, {Ade}, {Aghanim}, {Alves}, {Aniano}, {Armitage-Caplan}, {Arnaud},
  {Ashdown}, {Atrio-Barandela}, \& et~al.}]{pla14}
{Planck Collaboration}, {Abergel}, A., {Ade}, P.~A.~R., {et~al.} 2014, \aap,
  571, A11

\bibitem[{{Rakowski} {et~al.}(2011){Rakowski}, {Laming}, {Hwang}, {Eriksen},
  {Ghavamian}, \& {Hughes}}]{rlh11}
{Rakowski}, C.~E., {Laming}, J.~M., {Hwang}, U., {et~al.} 2011, \apjl, 735, L21

\bibitem[{{Raymond} {et~al.}(2007){Raymond}, {Korreck}, {Sedlacek}, {Blair},
  {Ghavamian}, \& {Sankrit}}]{rks07}
{Raymond}, J.~C., {Korreck}, K.~E., {Sedlacek}, Q.~C., {et~al.} 2007, \apj,
  659, 1257

\bibitem[{{Ressler} {et~al.}(2014){Ressler}, {Katsuda}, {Reynolds}, {Long},
  {Petre}, {Williams}, \& {Winkler}}]{rkr14}
{Ressler}, S.~M., {Katsuda}, S., {Reynolds}, S.~P., {et~al.} 2014, \apj, 790,
  85

\bibitem[{{Reynolds}(2008)}]{rey08}
{Reynolds}, S.~P. 2008, \araa, 46, 89

\bibitem[{{Reynoso} {et~al.}(2013){Reynoso}, {Hughes}, \& {Moffett}}]{rhm13}
{Reynoso}, E.~M., {Hughes}, J.~P., \& {Moffett}, D.~A. 2013, \aj, 145, 104

\bibitem[{{Rothenflug} {et~al.}(2004){Rothenflug}, {Ballet}, {Dubner},
  {Giacani}, {Decourchelle}, \& {Ferrando}}]{rbd04}
{Rothenflug}, R., {Ballet}, J., {Dubner}, G., {et~al.} 2004, \aap, 425, 121

\bibitem[{{Schure} {et~al.}(2012){Schure}, {Bell}, {O'C Drury}, \&
  {Bykov}}]{sbd12}
{Schure}, K.~M., {Bell}, A.~R., {O'C Drury}, L., \& {Bykov}, A.~M. 2012, \ssr,
  173, 491

\bibitem[{{Uchida} {et~al.}(2013){Uchida}, {Yamaguchi}, \& {Koyama}}]{uyk13}
{Uchida}, H., {Yamaguchi}, H., \& {Koyama}, K. 2013, \apj, 771, 56

\bibitem[{{Vink}(2012)}]{vin12}
{Vink}, J. 2012, \aapr, 20, 49

\bibitem[{{Winkler} {et~al.}(2003){Winkler}, {Gupta}, \& {Long}}]{wgl03}
{Winkler}, P.~F., {Gupta}, G., \& {Long}, K.~S. 2003, \apj, 585, 324

\bibitem[{{Winkler} {et~al.}(2013){Winkler}, {Williams}, {Blair}, {Borkowski},
  {Ghavamian}, {Long}, {Raymond}, \& {Reynolds}}]{wwb13}
{Winkler}, P.~F., {Williams}, B.~J., {Blair}, W.~P., {et~al.} 2013, \apj, 764,
  156

\bibitem[{{Winkler} {et~al.}(2014){Winkler}, {Williams}, {Reynolds}, {Petre},
  {Long}, {Katsuda}, \& {Hwang}}]{wwr14}
{Winkler}, P.~F., {Williams}, B.~J., {Reynolds}, S.~P., {et~al.} 2014, \apj,
  781, 65

\bibitem[{{Zirakashvili} \& {Aharonian}(2007)}]{za07}
{Zirakashvili}, V.~N. \& {Aharonian}, F. 2007, \aap, 465, 695

\end{thebibliography}


\end{document}